\documentclass[a4paper]{JHEP3} 

\usepackage{bbm,amsmath,graphicx,amssymb}
\usepackage{bm}
\usepackage{cite}
\usepackage{verbatim}
\usepackage{slashed}
\def\nn{\nonumber}
\def\tr{\mathop\textrm{tr}\nolimit}

\def\cA{\mathcal{A}}

\def\stamp{--- {\bf \today} --- {\bf \jobname.tex}}
\def\fss#1#2{\sin(\pi \alpha'\, #1\cdot #2)}
\def\fs_#1{\mathfrak{s}(#1)}
\def\cS{\mathcal{S}}
\def\tr{\textrm{tr}}

\def\cN{\mathcal{N}}

\def\BE{\begin{equation}}
\def\EE{\end{equation}}
\def\spa#1.#2{\left\langle#1\,#2\right\rangle}
\def\spb#1.#2{\left[#1\,#2\right]}
\def\lor#1.#2{\left(#1\,#2\right)}

\newcommand\fverb{\setbox\fverbbox=\hbox\bgroup\verb}
\newcommand\fverbdo{\egroup\medskip\noindent%
			\fbox{\unhbox\fverbbox}\ }
\newcommand\fverbit{\egroup\item[\fbox{\unhbox\fverbbox}]}
\newbox\fverbbox

\title{The\! Momentum\! Kernel\! of\! Gauge\! and\! Gravity\! Theories}

\author{N.E.J. Bjerrum-Bohr, Poul H. Damgaard and Thomas S{\o}ndergaard\\
Niels Bohr International Academy and Discovery Center,\\ The Niels Bohr
Institute,
Blegdamsvej 17,\\ DK-2100 Copenhagen \O, Denmark,\\ {\tt email:}
\{bjbohr;phdamg;tsonderg\}@nbi.dk}
\author{Pierre Vanhove\\
CEA, DSM, Institut de Physique Th{\'e}orique, IPhT, CNRS, MPPU,\\
URA2306, Saclay, F-91191 Gif-sur-Yvette, France,\\ {\tt email:}
pierre.vanhove@cea.fr}

\received{\today} 		
\accepted{\today}		

\preprint{IPHT-T10/145, IHES/P/10/45}
\abstract{
We derive an explicit formula for factorizing an $n$-point
closed string amplitude into open string amplitudes.
Our results are phrased in terms of a momentum kernel which in the
limit of infinite string tension reduces to the corresponding
field theory kernel. The same momentum kernel encodes the monodromy
relations which lead to the minimal basis of color-ordered
amplitudes in Yang-Mills theory. There are interesting
consequences of the momentum kernel pertaining to soft limits
of amplitudes. We also comment on surprising links between 
gravity and certain combinations of kinematic and color factors 
in gauge theory.}


\begin{document}
\section{Introduction}
Close to a century after its first formulation, the general
theory of relativity continues to give us surprises. In
particular, remarkable new insight arises from the intimate
relation between gravity and Yang-Mills theories. In
perturbation theory one of the most intriguing manifestations
of this comes from the factorization of closed string
amplitudes into products of open string amplitudes, the
Kawai-Lewellen-Tye (KLT) relations~\cite{Kawai:1985xq}. In the
field theory limit, these relations show that tree-level
amplitudes of gravitons can be expressed as products of
color-ordered Yang-Mills amplitudes. Recently, these KLT-relations
between gravity amplitudes and Yang-Mills amplitudes have been
proven in a series of different but equivalent forms using only
concepts from relativistic quantum field
theory~\cite{BjerrumBohr:2010ta,BjerrumBohr:2010yc}.
The explicit $n$-point formula that was conjectured
in ref.~\cite{Bern:1998sv} has been proven as one particular
case~\cite{BjerrumBohr:2010yc}.

One interesting by-product of the field theory derivation
provided in refs.~\cite{BjerrumBohr:2010ta,BjerrumBohr:2010yc}
was the observation that KLT-relations are closely linked to a
set of newly discovered identities among gauge theory
amplitudes, the BCJ-relations~\cite{Bern:2008qj}. These
BCJ-relations, which also hold when including
matter~\cite{Sondergaard:2009za,Jia:2010nz}, turn out to be the
field theory limit of monodromy relations in string
theory~\cite{BjerrumBohr:2009rd} (see
also~\cite{Stieberger:2009hq,Plahte:1970wy}). The field theory
proof of KLT-relations uses the same on-shell recursion
techniques as the field theory proof of
BCJ-relations~\cite{Feng:2010my}. But the connection between
the two sets of relations runs deeper. Indeed, monodromy
relations and hence BCJ-relations are implicitly implied by the
analysis of the original KLT paper~\cite{Kawai:1985xq}.

Although the recipe for deriving the $n$-point KLT-relations
between closed and open string amplitudes was given in
ref.~\cite{Kawai:1985xq}, a general explicit expression valid
for all $n$ has never been provided. In this paper we will derive
the explicit formulae for any $n$. A central object that
emerges is a momentum kernel
\begin{equation}
  \mathcal{S}_{\alpha'}[i_1,\ldots,i_k|
j_1,\ldots,j_k]_{p} \equiv (\pi\alpha'/2)^{-k}\,
\prod_{t=1}^{k}\, \sin \big(\pi\alpha'\,(p\cdot
  k_{i_t}+ \sum_{q>t}^{k} \, \theta(i_t,i_q)\, k_{i_t}\cdot k_{i_q})  \big)\,,
\end{equation}
whose precise definition will be provided below. This momentum
kernel maps products of open string amplitudes to closed string
amplitudes. In the field theory limit $\alpha' \to 0$ it turns
into the field theory momentum kernel ${\cal S}$ that maps
field theory Yang-Mills amplitudes to gravity
amplitudes~\cite{BjerrumBohr:2010ta,BjerrumBohr:2010yc}. In
beautiful agreement with the corresponding observation in field
theory, the string theory momentum kernel ${\cal S}_{\alpha'}$
is precisely the generator of monodromy relations. Phrased
more precisely, it annihilates color-ordered amplitudes
$\mathcal{A}_n$ according to
\begin{equation}
\sum_{\sigma} \mathcal{S}_{\alpha'}
[\sigma(2,\dots,n-1)|\beta(2,\dots,n-1)]_{k_1}
\mathcal{A}_n(n,\sigma(2,\dots,n-1),1) = 0\,,
\end{equation}
where $\beta$ is  any permutation of leg $2,\dots,n-1$  and the sum runs
over  all permutations  of these  legs.

Particularly interesting insight arises if one uses a
construction based on the heterotic string~\cite{Kawai:1985xq}.
This shows that Yang-Mills amplitudes $A_n^{YM}$ can be
expressed as a sum of products   of   color-ordered Yang-Mills
amplitudes $A_n$, $A_n^{YM}\sim   \sum  \,
A_n\,\mathcal{S}_{\alpha'}\,\widetilde{A}^{s}_n$, where
$\widetilde{A}^{s}_n$ is a {\em scalar} amplitude based on
vertices that are trivial except for the structure    constants
of the    gauge group~\cite{Bern:1999bx}. The similarity with
the relation for gravity amplitudes $M_n\sim  \sum A_n\,
\mathcal{S}_{\alpha'} \widetilde{A}_n$ in terms of gauge
amplitudes is striking. This leads naturally to an alternative
viewpoint on KLT-relations that comes from exploring
Jacobi-like identities~\cite{Bern:2008qj} among numerator
factors on the Yang-Mills side. Also this picture, and its
generalization to extended Jacobi-like structures, has a
natural string theory
interpretation~\cite{Mafra:2009bz,Tye:2010dd,BjerrumBohr:2010zs}.
This becomes particularly transparent within the framework of
the heterotic string~\cite{Tye:2010dd} (see
also~\cite{Cheung:2010vn,Boels:2010bv}). There is hope that
much of this tree-level
structure~\cite{Bern:2010yg,Vaman:2010ez} carries over to any
number of loops~\cite{Bern:2010ue}. This can be used to analyze
the ultraviolet  behaviour  of $\cN=8$  supergravity~\cite{Vanhove:2010nf}. 
All of this is strong motivation
for analyzing the numerator approach and the associated duality
between kinematic and color structures in greater detail.

Finally, an interesting case arises when one applies the field
theory momentum kernel ${\cal S}$ to amplitudes with mismatched
external helicity legs. Then one obtains new non-linear
relations among Yang-Mills
amplitudes~\cite{BjerrumBohr:2010zb} that can be understood in the context of
$R$-charges~\cite{Tye:2010kg,Feng:2010br,Feng:2010hd,Elvang:2010kc}.

This paper will be devoted to a detailed study of the string
theory momentum kernel ${\cal S}_{\alpha'}$. First we derive,
for the first time, the explicit $n$-point relations between
closed string and open string amplitudes, and thus demonstrate
in detail how the momentum kernel arises in string theory. We
next explore some of the properties of this momentum kernel,
and in particular we explain how it acts as a generator of
monodromy relations. It turns out that there is, just as in
field theory \cite{BjerrumBohr:2010zb}, a great amount of
freedom in writing down the explicit KLT-map, a freedom which
is directly related to this monodromy. Common to {\em all}
rewritings is, however, the same momentum kernel ${\cal
S}_{\alpha'}$. In the last part of the paper we make some
observations regarding the momentum kernel ${\cal S}$ in the
field theory limit and soft factorization of amplitudes in
gauge theory and gravity. We also comment on the relation to
the approach based on Jacobi-like structures among amplitudes.

\section{The momentum kernel in string theory\label{stringkernel}}

In this section we will derive a general form of
$n$-point closed string amplitudes in terms of products of
color-ordered open string amplitudes. Like in the  KLT
paper~\cite{Kawai:1985xq}, we  proceed  by explicit holomorphic
factorization of a closed string amplitude.

The heterotic and open strings have different spectra that
leads to different effective actions. In the field theory
$\alpha'\to 0$ limit the heterotic and open string tree-level
amplitudes reduce to various gauge or gravity amplitudes. 
Closed type~II and heterotic string tree-level gravity
amplitudes are different because of the different spectra in
the two theories. $\cN=8$ supergravity amplitudes are computed
as the field theory limit of closed string type~II amplitudes,
and $\cN=4$ supergravity amplitudes as the field theory limit
of heterotic string amplitudes. For multi-graviton amplitudes
the difference arises at higher order in $\alpha'$ and it does not
affect the field theory limit.

When performing the holomorphic factorization of the amplitudes
we will not have to specify the actual detailed form of the
left and right moving string amplitudes. But we will pay attention to
the monodromy properties of the amplitude when the positions of
the external states move on the sphere. All the monodromy
properties~\cite{BjerrumBohr:2009rd,BjerrumBohr:2010zs} arise
from the contraction between the corresponding plane-wave
factors. In this way universal relations emerge.

\subsection{Gravity and gauge theory amplitudes}\label{sec:s-factor}
After fixing the three points $z_1=0$,
$z_{n-1}=1$ and  $z_n=\infty$, the $n$-point closed string
amplitude takes the general form
\begin{equation}
  \label{eq:DefMn}
  \mathcal{M}_n=\left(i\over 2\pi\alpha'\right)^{n-3}\,
\!\!\!\!\int \prod_{i=2}^{n-2} d^2z_i |z_i|^{2\alpha' k_1\cdot k_i} |z_i-1|^{2\alpha'\,k_{n-1}\cdot k_i}\!\!\!\!
\prod_{i<j\leq
    n-2}\!\! |z_j-z_i|^{2\alpha'\,k_i\cdot k_j}\, f(z_i)\,g(\bar z_i)\, ,
\end{equation}
where $f(z_i)$ and $g(\bar  z_i)$ arise from the operator
product expansion of the vertex operators. They are functions
without branch cuts. The precise form of these functions
depends on the external states. They can be gravitons or gauge
fields, or, in the straightforward supersymmetric
generalization, any part of the ${\cal N}=4$ supermultiplet.
However, the exact form of these functions will not affect the
general relations. For definiteness, let us from now on
consider ${\cal M}_n$ to be a gravity amplitude.

Following~\cite{Dotsenko:1984nm,Dotsenko:1984ad} we can write
eq.~\eqref{eq:DefMn} in terms of $v_i^1$ and $v_i^2$, where
$z_i = v_i^1 +iv_i^2$, and then make the following change of variables
for $v_i^2$
\begin{align}
v_i^2 \quad \longrightarrow \quad ie^{-2i\epsilon}v_i^2,
\end{align}
where $\epsilon>0$ is some small number. Keeping  only the  terms
linear in $\epsilon$, \textit{i.e.} using 
\begin{equation}
ie^{-2i\epsilon}v_i^2 \simeq i(1-2i\epsilon)v_i^2\,,
\end{equation} 
and introducing the notation
\begin{align}
v_i^\pm \equiv  v_i^1\pm v_i^2 \,, \qquad \delta_i \equiv v_i^+-v_i^-\,,
\end{align}
we can write eq.~\eqref{eq:DefMn} as an  `almost
factorized' amplitude
\begin{align}
\label{eq:DefMn1}
&\mathcal{M}_n= \left(\frac{i}{2}\right)^{n-3}\left(i\over 2\pi\alpha'\right)^{n-3}\!\!
\int_{-\infty}^{+\infty} \prod_{i=2}^{n-2} dv_i^+dv_i^-
f(v^-_i)\,g(v_i^+)\nonumber \\
&\hspace{1cm}\times 
(v_i^+-i\epsilon\delta_i)^{\alpha' k_1\cdot k_i}
(v_i^-+i\epsilon\delta_i)^{\alpha' k_1\cdot k_i}
(v_i^+-1-i\epsilon\delta_i)^{\alpha'\,k_{n-1}\cdot k_i}
(v_i^--1+i\epsilon\delta_i)^{\alpha'\,k_{n-1}\cdot k_i}\nonumber \\
&\hspace{1cm}\times \prod_{i<j\leq n-2} \big(v_i^+-v_j^+
-i\epsilon (\delta_i-\delta_j)\big)^{\alpha'\,k_i\cdot k_j}
\big(v_i^--v_j^-+i\epsilon (\delta_i-\delta_j)\big)^{\alpha'\,k_i\cdot k_j}\,.
\end{align}

The appearance of branch cuts in the integrand is crucial.
We have the following choices for $x^\alpha$ when $x<0$
\begin{equation}
x^\alpha= \begin{cases}
e^{i\pi\alpha} (-x)^\alpha& {\rm Im} (x)\geq 0\,,\cr
e^{-i\pi\alpha}\, (-x)^\alpha& {\rm Im} (x)<0\,.
\end{cases}
\end{equation}
By splitting the $v_i^+$-integrals and using symmetry, the $n$-point graviton amplitude
can be written as
\begin{equation}\label{e:Msum}
\mathcal{M}_n=\sum_{\sigma}          \,
M_n\big(\sigma(2),\ldots,\sigma(n-2)\big)\,,
\end{equation}
where $M_n(\sigma(2),\cdots,\sigma(n-2))$ is the ordered
amplitude defined such that $v^+_{\sigma(2)}    <
v^+_{\sigma(3)}<   \cdots    < v^+_{\sigma(n-2)}$.

If one of the $v^+_i$ is in the interval $]-\infty,0[$ the
integral contours for $v^-_i$ lie below the real axis and the
integral vanishes. This is because there are no poles of the
functions $f$  and $g$ outside the real axis, and the integrand
nicely vanishes at infinity. By the same reasoning, if one of
the $v^+_i$ is in the interval $]1,+\infty[$, the integral
contours for $v^-_i$ lie above the real axis and the integral
is again vanishing.

Therefore, to get a non-vanishing contribution, all the $v_+$'s
need to be distributed in the interval $]0,1[$:
$0<v^+_{\sigma(2)} < v^+_{\sigma(3)}< \cdots <
v^+_{\sigma(n-2)}<1$ where $\sigma$ denotes a permutation of
the $(n-3)!$ labels.  These $v^+_i$  integrals lead to
\begin{align}
  \cA_n(1,\sigma(2,\dots,n-2),n-1,n)={}& \int_{ 0<v^+_{\sigma(2)} <  \cdots <
v^+_{\sigma(n-2)}<1}       \prod_{i=2}^{n-2}        dv_{\sigma(i)}^+\,
  g\big(v^+_{\sigma(i)}\big)\nonumber \\
&\hspace{-4.5cm}\times
  \big(v_{\sigma(i)}^+\big)^{\alpha' k_1\cdot k_{\sigma(i)}}
\big(1-v_{\sigma(i)}^+\big)^{\alpha'\,k_{n-1}\cdot k_{\sigma(i)}}
\prod_{{\sigma(i)}<{\sigma(j)}\leq
    n-2}\big(v_{\sigma(j)}^+-v_{\sigma(i)}^+\big)^{\alpha'\,k_{\sigma(i)}\cdot k_{\sigma(j)}}\,,
\label{e:ALfinal}\end{align}
which are integrals corresponding to color-ordered open string
amplitudes.

We  now turn  to  the evaluation  of  the integrals  over the
$v^-_i$ variables.  This  will  lead  to  ordered  integrals
$\widetilde{\cA}_n$  that correspond to  the `right-moving'
sectors. For simplicity we just  consider the  case of ordering
$\{2,3,\dots ,n-2\}$  of the $v_i^+$ variables.  All other
cases  are obtained by summing  over the permutations of these
$n-3$ variables.

\begin{figure}[t]
\centering
\includegraphics[width=12cm]{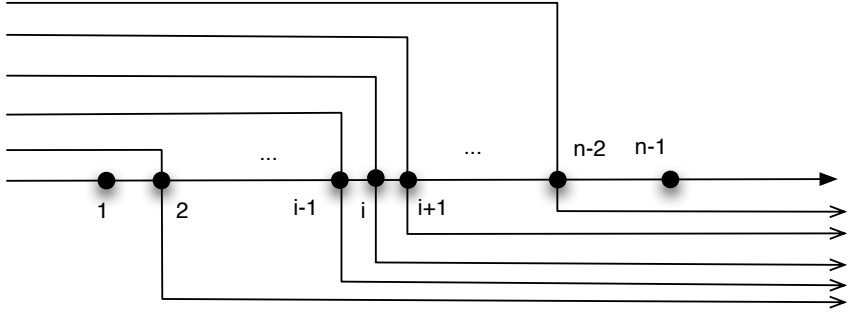}
\caption{\sl The nested structure of the contours of integration
for the variable $v^-_i$   corresponding       to        the       ordering
$0<v^+_2<v^+_3<\cdots<v^+_{n-2}<1$ of the $v_+$ variables.}
\label{fig:contourNested}
\end{figure}

For each $2\leq i\leq n-2$ we first examine the behavior of the
integrand of~\eqref{eq:DefMn} around the different branch cuts.
For $v_i^-\sim 0$ the quantity $v^-_i+i\epsilon \delta_i\sim
v^-_i+i\epsilon v_i^+$ has a positive imaginary part. Therefore
the contour is above the real axis. For $v_i^-\sim 1$ the
quantity $v^-_i-1+i\epsilon \delta_i\sim v^-_i-1+i\epsilon
(v_i^+-1)$ has a negative imaginary part; the contour of
integration lies below the real axis.  Finally, for $v^-_i\sim
v^-_j$ with $i<j$, the quantity $v^-_i-v^-_j+i\epsilon
(\delta_i-\delta_j)\sim v^-_i-v^-_j+i\epsilon (v_i^+-v_j^+)$
has a negative imaginary part. Therefore the contour of
integration for $v^-_i$ goes below the contour of $v^-_j$ for
$i<j$.  We have represented this nested structure of the
contours of integration for the $v^-_i$ variables in
figure~\ref{fig:contourNested}.

We now consider the deformations of the contours of integration
for the $v^-_i$ variables.  Because the contours cannot cross
each other we need to close them either to the right, turning
around the branch cut at $z=1$ by starting with the rightmost,
or close the contours to the left, turning around the branch
cut at $z=0$, starting with the leftmost.

There is evidently an arbitrariness in the number of contours
that are closed to the left or closed to the right. For a given
$2\leq j\leq n-2$, we can pull the contours for the set between
2 and $j-1$ to the left, and the set between $j$ and $n-2$ to
the right.

Pulling the contour for 2 gives
\begin{align}
  \label{e:C2}
  &\int_{C_2}          dv_2^-  \,       (v_2^-)^{\alpha'k_1\cdot         k_2}
  (1-v_2^-)^{\alpha'k_{n-1}\cdot                  k_2}\,\prod_{j=3}^{n-2}
  (v_j^--v_2^-)^{\alpha'k_j\cdot k_2}\,f(v_2^-) \nonumber \\
&=2i\sin(\pi \alpha' k_1\cdot k_2)
  \int_{-\infty}^0 dv_2^-\,        (-v_2^-)^{\alpha'k_1\cdot         k_2}
  (1-v_2^-)^{\alpha'k_{n-1}\cdot                  k_2}\,\prod_{j=3}^{n-2}
  (v_j^--v_2^-)^{\alpha'k_j\cdot k_2}\,f(v_2^-)\,.
\end{align}
Here we  have explicitly shown only  the contributions where
$v_2^-$ has branch cuts.

Closing the contour for $v_3^-$ to the left leads to
\begin{align}
  \label{e:C3}
\nn  &\!\!\!\!\!\!\int_{C_3}          dv_3^-   \,      (v_3^-)^{\alpha'k_1\cdot k_3}
  (1-v_3^-)^{\alpha'k_{n-1}\cdot k_3}(v_3^--v_2^-)^{\alpha'k_2\cdot k_3}\,\prod_{j=4}^{n-2}
  (v_j^--v_3^-)^{\alpha'k_j\cdot k_3}\,f(v_3^-)\\
&\!\!\!\!\!\!=2i\sin(\pi \alpha' k_1\cdot k_3)
  \int_{v_2^-}^{0} dv_3^-\,     (-v_3^-)^{\alpha'k_1\cdot         k_3}
  (1-v_3^-)^{\alpha'k_{n-1}\cdot k_3}(v_3^--v_2^-)^{\alpha'k_2\cdot k_3}\\
\nn&\hspace{9cm}\times \prod_{j=4}^{n-2}
  (v_j^--v_3^-)^{\alpha'k_j\cdot k_3}\,f(v_3^-)\\
\nn&\!\!\!\!\!\! +2i\sin\big(\pi \alpha' (k_1+k_2)\cdot k_3\big)  \,  \int_{-\infty}^{v_2^-} dv_3^-\,
(-v_3^-)^{\alpha'k_1\cdot         k_3}
  (1-v_3^-)^{\alpha'k_{n-1}\cdot k_3}(v_2^--v_3^-)^{\alpha'k_2\cdot k_3}\\
\nn&\hspace{9cm}\times \prod_{j=4}^{n-2}
  (v_j^--v_3^-)^{\alpha'k_j\cdot k_3}\,f(v_3^-)\,,
\end{align}
and so on until one has pulled the contour for $v_{j-1}^-$ to
the left.

When closing  the contours to the  right we start from the
contour for $v_{n-2}^-$ down to the one for $v_j^-$. Pulling
first the contour for $v_{n-2}^-$ to the right leads to
\begin{align}
  \label{e:Cn-2}
  &\int_{C_{n-2}}          dv_{n-2}^-   \,
(v^-_{n-2})^{\alpha'k_1\cdot         k_{n-2}}
  (1-v^-_{n-2})^{\alpha'k_{n-1}\cdot
k_{n-2}}\,\prod_{j=2}^{n-3}
  (v_{n-2}^--v_j^-)^{\alpha'k_j\cdot k_{n-2}}\,f(v_{n-2}^-)\cr
&=2i\sin(\pi \alpha' k_{n-1}\cdot k_{n-2})
  \int_{1}^{+\infty} dv_{n-2}^-\,
(v^-_{n-2})^{\alpha'k_1\cdot         k_{n-2}}
  (v_{n-2}^--1)^{\alpha'k_{n-1}\cdot
k_{n-2}}\\
\nn&\hspace{9cm}\times \,\prod_{j=2}^{n-3}
  (v_{n-2}^--v_j^-)^{\alpha'k_j\cdot k_{n-2}}\,f(v_{n-2}^-)\,.
\end{align}
Similarly, closing the contour for $v_{n-3}^-$ to the right
leads to
\begin{align}
  \label{e:Cn-3}
 \nn &\int_{C_{n-3}}          dv_{n-3}^-   \,
(v_3^-)^{\alpha'k_1\cdot         k_{n-3}}
  (1-v_{n-3}^-)^{\alpha'k_{n-1}\cdot k_{n-3}}(v_{n-2}^--v_{n-3}^-)^{\alpha'k_{n-2}\cdot k_{n-3}}\\
\nn&\hspace{9cm}\times \prod_{j=2}^{n-4}
  (v_{n-3}^--v_j^-)^{\alpha'k_j\cdot k_{n-3}}\,f(v_{n-3}^-)\\
&=2i\sin(\pi \alpha' k_{n-1}\cdot k_{n-3})
  \int_{1}^{v_{n-2}^-} dv_{n-3}^-\,(v_{n-3}^-)^{\alpha'k_1\cdot k_{n-3}}
  (v_{n-3}^--1)^{\alpha'k_{n-1}\cdot k_{n-3}}\,\\ & \nonumber\hspace{5cm}\times
(v_{n-2}^--v_{n-3}^-)^{\alpha'k_{n-2}\cdot k_{n-3}}\,\prod_{j=2}^{n-4}
  (v_{n-3}^--v_j^-)^{\alpha'k_j\cdot k_{n-3}}\,f(v_{n-3}^-)\\
\nn&+2i\sin\big(\pi \alpha' (k_{n-1}+k_{n-2})\cdot k_{n-3}\big) \,
\int_{v_{n-2}^-}^{+\infty} dv_{n-3}^-\,
(v_{n-3}^-)^{\alpha'k_{n-1}\cdot         k_{n-3}}
  (v_{n-3}^--1)^{\alpha'k_{n-1}\cdot k_{n-3}}\,\\ & \nonumber\hspace{5cm}\times
(v_{n-3}^--v_{n-2}^-)^{\alpha'k_{n-2}\cdot k_{n-3}}\,\prod_{j=2}^{n-4}
  (v_{n-3}^--v_j^-)^{\alpha'k_j\cdot k_{n-3}}\,f(v_{n-3}^-)\,,
\end{align}
and so on until one reaches the contour for $v_j^-$. The
integrals  over the $v^-$  variables are the  ordered string
amplitudes
$\widetilde{\cA}_n(\gamma(2,\dots,j-1),1,n-1,\beta(j,\dots,n-2),n)$.

Collecting these contour deformations lead to the following
expression for  the  $v^-$  part  of  the integral
in~(\ref{eq:DefMn})
\begin{align}
&\!\left(-i/4\right)^{n-3}\sum_{\gamma}
\sum_{\beta}\,
 \mathcal{S}_{\alpha'}[\gamma(2,\dots,j-1)|2,\dots,j-1]_{k_1}\,
  \mathcal{S}_{\alpha'}[\beta(j,\dots,n-2)|j,\dots,n-2]_{k_{n-1}} \nonumber \\
&\hspace{4.5cm}\times
\label{e:ARfinal} \widetilde{\cA}_n(\gamma(2,\dots,j-1),1,n-1,\beta(j,\dots,n-2),n)\,,
\end{align}
which is of course multiplied by the left-moving amplitude
$\cA_n(1,2,\dots,n)$ from the integral over the  $v^+$
variables. Here we also see the first appearance of the
\textit{momentum kernel}
$\mathcal{S}_{\alpha'}[\gamma,\sigma]_p$. It depends on the
permutations $\gamma$ of  the $v_i^-$-variables and the
ordering  of the $v_i^+$. It also depends on the momenta
$p=k_1$ and $k_{n-1}$ of the  states at the  branch cut at $z=
0$  or $z= 1$  on the  sphere. In general it can be defined as
\begin{equation}
  \label{e:Cgs}
  \mathcal{S}_{\alpha'}[i_1,\ldots,i_k|
j_1,\ldots,j_k]_{p} \equiv (\pi\alpha'/2)^{-k}\,
\prod_{t=1}^{k}\, \sin \big(\pi\alpha'\,(p\cdot
  k_{i_t}+ \sum_{q>t}^{k} \, \theta(i_t,i_q)\, k_{i_t}\cdot k_{i_q})  \big)\,,
\end{equation}
where $\theta(i_t,i_q)$ equals 1 if the ordering of the legs
$i_t$ and $i_q$ is opposite in the sets $\{i_1,\ldots,i_k\}$
and $\{j_1,\ldots,j_k\}$, and 0 if the ordering is the same.

We have normalized this expression so that  in the  field
theory limit $\alpha'\rightarrow 0$ the kernel
$\mathcal{S}_{\alpha'}$ reduces to the field theory kernel
$\mathcal{S}$
of~\cite{BjerrumBohr:2010ta,BjerrumBohr:2010zb,BjerrumBohr:2010yc}.
As indicated, we will distinguish between these  two momentum
kernels by putting a subscript $\alpha'$ on the one of string
theory. In the next section we will analyze the properties of
this kernel.

To get  the full closed  string amplitude~(\ref{eq:DefMn}) we
need to multiply the left-moving amplitude of the $v^+$
integrations in~\eqref{e:ALfinal} with the right-moving
contribution in~\eqref{e:ARfinal} and then sum over all
orderings to get
\begin{align}
  \label{eq:Mnfinal}\nonumber
 \mathcal{M}_n ={}& \left(-i/4\right)^{n-3}\times\\&\hspace{-1.2cm}\sum_{\sigma}
\sum_{\gamma,\beta}
\mathcal{S}_{\alpha'}[\gamma(\sigma(2),\dots,\sigma(j\!-\!1))|\sigma(2,\dots,j\!-\!1)]_{k_1}
\mathcal{S}_{\alpha'}[\beta(\sigma(j),\dots,\sigma(n\!-\! 2))|
\sigma(j,\dots,n\!-\! 2)]_{k_{n\!-\!1}}\nonumber \\
&\hspace{-0.9cm} \times
   \cA_n(1,\sigma(2,\dots,n\!-\! 2),n\!-\! 1,n)\,
   \widetilde{\cA}_n(\gamma(\sigma(2),\dots,\sigma(j\!-\! 1)),1,n\!-\!1,\beta(\sigma(j),\dots,\sigma(n\!-\!2)),n)\,.
\end{align}
This provide a general form of the closed/open string
relation between external gauge bosons and gravitons at
tree-level. When restricted to the case of graviton external
states the field theory limit of this expression reduces to the
form derived in~\cite{BjerrumBohr:2010zb,BjerrumBohr:2010yc}.

As seen from the above derivation,
expression~\eqref{eq:Mnfinal} is actually independent of the
value of $j$. This $j$-independence reflects the arbitrariness
in the number of contours one closes to the left around the
branch point at $z=0$ or to the right around the branch point
at $z=1$. As further explained below, independence under shifts
of $j$ is a consequence of the monodromy
relations~\cite{BjerrumBohr:2009rd} that are satisfied by  the
color-ordered string amplitudes.

The expression~\eqref{eq:Mnfinal} is a sum over $(n-3)!\times
(j-2)!\times (n-1-j)!$
 terms.  The number of terms takes the maximal value
$(n-3)!\times (n-3)!$ for  $j=2$  or  $j=n-1$.  The choice made by
KLT in~\cite{Kawai:1985xq} consists in closing
half of the contours to the left and and the other
half to the right, i.e. $j=\lceil n/2\rceil$.  This leads to the smallest number of
terms  
$  (n-3)!\times(\left\lceil {n\over2}
\right\rceil -2)!\times (\left\lfloor {n\over2}\right\rfloor-1)!
$
(The  floor  and  ceiling
  functions are defined on half-integers as follows:
  $\lfloor n/2\rfloor=(n-1)/2$ if $n$ is odd, or $n/2$ if
$n$ is even. $\lceil n/2\rceil=(n+1)/2$ if $n$ is odd, or $n/2$ if
$n$ is even.)\\
For $j=n-1$ the amplitude takes the nice form
\begin{align}
\label{stringPureKLTn}
\mathcal{M}_n &=(-1)^{n-3} \sum_{\sigma,\gamma}
\mathcal{S}_{\alpha'}[\gamma(2,\dots,n-2)|\sigma(2,\dots,n-2)]_{k_1}\cr
&\hspace{1cm}\times \cA_n(1,\sigma(2,\dots,n-2),n-1,n)
\widetilde{\cA}_n(n-1,n,\gamma(2,\dots,n-2),1)\,.
\end{align}
%

\section{Properties of the momentum kernel \label{s:genpropS}}

The $\mathcal{S}_{\alpha'}$ kernel has a number of fundamental
properties that correspond to those of the field theory kernel
$\mathcal{S}$. In field theory these properties ensure, for
instance, the correct cancellation of poles between products of
color-ordered gauge amplitudes and correct factorization
properties compatible with on-shell recursion techniques. The
$\mathcal{S}_{\alpha'}$ version of these properties can be seen
to hold by the same kind of arguments used for the field
theory $\mathcal{S}$ kernel
in~\cite{BjerrumBohr:2010ta,BjerrumBohr:2010zb,BjerrumBohr:2010yc},
and will therefore not be repeated here.

\medskip
\medskip\noindent {\sl (1) Reflection symmetry:}
\begin{equation}\label{e:s1}
\mathcal{S}_{\alpha'}[\sigma(1,\dots,k)|\gamma(1,\dots,k)]_p =
\mathcal{S}_{\alpha'}[\gamma(k,\dots,1)|\sigma(k,\dots,1)]_{p}\,,
\end{equation}
where $p$ is a massless momentum and $\sigma$ and $\gamma$ are arbitrary
permutations of the $k$ labels $\{1,\dots,k\}$.

\medskip
\medskip\noindent{\sl (2) Factorization:}
Assuming that $P \equiv k_1 + k_2 +\cdots + k_p$ is on-shell,
\textit{i.e.} $P^2 = 0$, the following factorization holds
\begin{equation}\label{e:fact}\begin{split}
&\mathcal{S}_{\alpha'}[\gamma(p+1,\dots,k)\sigma(2,\dots,p)|
\beta(2,\dots,p)\delta(p+1,\dots,k)]_{k_1} \cr
&= \mathcal{S}_{\alpha'}[\sigma(2,\dots,p)|\beta(2,\dots,p)]_{k_1} \times
\mathcal{S}_{\alpha'}[\gamma(p+1,\dots,k)|\delta(p+1,\dots,k)]_{P}\,,
\end{split}\end{equation}
where $\alpha, \beta, \gamma$ and $\delta$ are arbitrary
permutations.

\medskip
\medskip\noindent{\sl (3) Annihilation of amplitudes:}
\begin{equation}
\sum_{\sigma} \mathcal{S}_{\alpha'}
[\sigma(2,\dots,n-1)|\beta(2,\dots,n-1)]_{k_1}
\mathcal{A}_n(n,\sigma(2,\dots,n-1),1) = 0\,,
\label{monoS}
\end{equation}
where $\beta$ is any permutation of the legs $\{ 2,\dots,n-1\}$
and    $\mathcal{A}_n$    are    color-ordered tree-level
string amplitudes.

\medskip
\medskip\noindent{\sl (4) The shifting-formula for $j$:}
By using  that formula~(\ref{eq:Mnfinal})  is
independent of  $j$ we  obtain the following  relation, which is
valid  for any $2\leq j\leq n-2$:

\begin{align}
\nn&\sum_{\gamma,\beta}
\mathcal{S}_{\alpha'}[\gamma(i_2,\dots,i_j)|i_2,\ldots,i_j]_{k_1}
\mathcal{S}_{\alpha'}[i_{j+1},\ldots,i_{n-2}|
\beta(i_{j+1},\dots,i_{n-2})]_{k_{n-1}}\\
&\hspace{2cm}\times
\mathcal{A}_n(\gamma(i_2,\dots,i_j),1,n-1,\beta(i_{j+1},
\dots,i_{n-2}),n) \nonumber \\
&=\sum_{\gamma',\beta'}
\mathcal{S}_{\alpha'}[\gamma'(i_2,\dots,i_{j-1})|i_2,\ldots,i_{j-1}]_{k_1}
\mathcal{S}_{\alpha'}[i_j,\ldots,i_{n-2}|
\beta'(i_j,\dots,i_{n-2})]_{k_{n-1}}\\
\nn&\hspace{2cm}\times
 \mathcal{A}_n(\gamma'(i_2,\dots,i_{j-1}),1,n-1,
\beta'(i_j,\dots,i_{n-2}),n)\,.
\end{align}
They are particular cases of linear monodromy relations
satisfied by the color-ordered amplitudes. Conversely,
monodromy relations are necessary and sufficient for proving
the $j$-independence of the general KLT
formula~\eqref{eq:Mnfinal}. Such monodromy properties are
generated by the contour deformations as discussed in
section~\ref{sec:s-factor}.

As mentioned, property (1)--(4) are also satisfied in field
theory simply by replacing $\mathcal{S}_{\alpha'}$ by its field
theory limit $\mathcal{S}$ and replacing the color-ordered
string amplitudes $\mathcal{A}_n$ by their corresponding field
theory limits $A_n$.

\subsection{Minimal basis from momentum kernel}\label{sec:minbas}
One can write the equations  in~(\ref{monoS}) as a linear
system  of the $(n-2)!-(n-3)!=(n-3)\times              (n-3)!$
amplitudes
$\cA_n(n,\sigma(2),\dots,\sigma(i-1),n-1,\sigma(i),\dots,\sigma(n-2),1)$
with  a  right-hand-side  expressed only  in a  minimal basis
of  size  $(n-3)!$,   spanned  by   the  amplitudes
$\cA_n(n,\sigma(2),\dots,\sigma(n-2),n-1,1)$
\begin{eqnarray}
  \nn
  &&\!\!\!\!\!  \sum_{i=2}^{n-2}\!\sum_{\sigma}
\cS_{\alpha'}[\sigma(2,\dots,n\!-\!1)|\beta(2,\dots,n\!-\!1)]_{k_1}
\cA_n(n,\sigma(2),\dots,\sigma(i\!-\!1),n\!-\!1,\sigma(i),\dots,\sigma(n\!-\!2),1) \\
  &&\!\!\!\!\!=-\sum_{\sigma}\,\cS_{\alpha'}
[\sigma(2,\dots,n\!-\!2),n\!-\!1|\beta(2,\dots,n\!-\!1)]_{k_1}\,
\cA_n(n,\sigma(2),\dots,\sigma(n\!-\!2),n\!-\!1,1) \,.
  \label{e:split3}
\end{eqnarray}
The  resulting  linear   system  is non-degenerate for a
generic choice  of momenta and can be inverted to express   the
amplitudes
$\cA_n(n,\sigma(2),\dots,\sigma(i-1),n-1,\sigma(i),\dots,\sigma(n-2),1)$
in       the       basis        of
$\cA_n(n,\sigma(2),\dots,\sigma(n-2),n-1,1)$.  Equivalently,
the rank of the  square matrix of size $(n-2)!$, with entries
$S[\sigma|\beta]$, has a kernel of dimension $(n-3)!$ and  the
linear system for the amplitudes
$\cA_n(n,\sigma(2),\dots,\sigma(i-1),n-1,\sigma(i),\dots,\sigma(n-2),1)$
in~(\ref{e:split3}) is $(n-3)\times (n-3)!$.  We have
numerically check this fact up to 10 points.

Below  we work  out  some specific  examples.   Especially, we
will illustrate  how the  monodromy  relations  with leg  $n$
and 1  fixed as in~(\ref{monoS}) are  enough~\cite{Feng:2010my}
to deduce  the minimal basis~\cite{BjerrumBohr:2009rd}.

At four points the momentum kernel takes the form (discarding
the overall normalization constant)
\begin{eqnarray}
\nn  \cS_{\alpha'}[23|23]_{k_1}&=&\cS_{\alpha'}
[32|32]_{k_1}=\fss{k_1}{k_2}\,\fss{k_1}{k_3},\\
\label{e:S4pt}\cS_{\alpha'}[23|32]_{k_1}&=&
-\fss{k_1}{k_3}^2,\qquad
\cS_{\alpha'}[32|23]_{k_1}=-\fss{k_1}{k_2}^2\,,
\end{eqnarray}
and the system of equations~\eqref{e:split3} gives
for $\beta(2,3) = \{2,3\}$ and  $\beta(2,3) = \{3,2\}$
\begin{eqnarray}
\nn\fss{k_1}{k_2}\fss{k_1}{k_3} \,\cA_4(4,2,3,1)
- \fss{k_1}{k_2}^2 \,\cA_4(4,3,2,1) &=& 0,\\
\hspace{-1.5cm}-\fss{k_1}{k_3}^2\,\cA_4(4,2,3,1) +
\fss{k_1}{k_2}\fss{k_1}{k_3}\,\cA_4(4,3,2,1) &=& 0\,.\ \ \ \ \ 
\end{eqnarray}
These two equations are identical and lead to the monodromy
relation
\begin{equation}
\fss{k_1}{k_3} \,\cA_4(4,2,3,1) =
 \fss{k_1}{k_2} \,\cA_4(4,3,2,1)  \,,
\end{equation}
reducing to a minimal basis.

The system of equations~\eqref{e:split3}
for $\beta(2,3,4) = \{2,3,4\}$ and  $\beta(2,3,4) = \{ 3,2,4\}$ implies that
\begin{equation}\label{e:sys5}\begin{split}\hspace{-0.5cm}\!\!
\fss{k_1}{k_2}\fss{k_1}{k_3} F(4\{23\}) \!+\!
\fss{k_1}{k_2} \fss{k_3}{(k_1\!+\!k_2)}F(4\{32\})\ &\!=\!0, \ \ \ \cr
 \fss{k_1}{k_3}\fss{k_2}{(k_1\!+\!k_3)}\,F(4\{23\})\!+\!
\fss{k_1}{k_2}\fss{k_1}{k_3}\, F(4\{32\})&\!=\!0\,,\ \ \ 
\end{split}
\end{equation}
where we have defined
\begin{eqnarray}
\nn F(4\{23\}) &\equiv&-\fss{k_4}{k_5}\,
\cA_5(5,4,2,3,1) + \fss{k_4}{(k_1+k_3)}\,\cA_5(5,2,4,3,1)\\
& &+ \fss{k_1}{k_4}\,\cA_5(5,2,3,4,1),  \nonumber \\
\nn F(4\{32\})&\equiv& -\fss{k_4}{k_5}\,
\cA_5(5,4,3,2,1) + \fss{k_4}{(k_1+k_2)}\,\cA_5(5,3,4,2,1)\\
& &+ \fss{k_1}{k_4}\,\cA_5(5,3,2,4,1)\,.
\label{fundone}
\end{eqnarray}
For generic values of the external momenta the determinant of
this system $\fss{k_1}{k_2}\fss{k_1}{k_3}
\fss{k_2}{k_3}\fss{k_4}{k_5}$ is non vanishing, implying that
$F(4\{23\})=0$ and $F(4\{32\})=0$. These identities are two of
the six monodromy relations~\cite{BjerrumBohr:2009rd} that one
obtains by keeping four legs fixed and rotating the contour of
one external leg (here leg 2). We have seen here how these
relations arise from the annihilation property~(3).

These two relations directly express the color-ordered
amplitudes $\cA_5(5,3,4,2,1)$ and $\cA_5(5,2,3,4,1)$ in the
minimal basis $\cA_5(1,2,3,4,5)$ and $\cA_5(1,3,2,4,5)$. By
considering all the other permutations of the external legs it
is straightforward to see that property~(3) implies that all
color-ordered amplitudes can be expressed in the  minimal basis
$\cA_5(1,2,3,4,5)$ and $\cA_5(1,3,2,4,5)$. The generalization
to higher $n$-point cases follows analogously.

\section{Soft limit of graviton amplitudes at tree-level}\label{sec:soft}
In this and in the following section we make some observations on 
the field theory limit.

Interestingly, tree-level gravity amplitudes have a universal
behavior when taking one graviton to be soft, a classic result
due to Weinberg~\cite{Weinberg:1965nx,Bern:1998sv}
\begin{align}
\lim_{k_s^\pm\to0} M_n(\ldots,a,s^{\pm},b,\ldots)=
S^{\mathrm{gravity}}(s^{\pm}) \times M_{n-1}(\ldots,a,b,\ldots)\,.
\label{softM}
\end{align}
For definiteness take leg $n$ to be soft, then the `soft
factor' is given by the sum
\begin{equation}
S^{\mathrm{gravity}}(n^\pm)  =\sum_{i=1}^{n-1}  \,  s_{n\,  i}\,S^{\rm
YM}(q_L,n^\pm,i) S^{\rm YM}(q_R,n^\pm,i)\,,
\label{softSM}
\end{equation}
where $S^{\rm YM}(q_L,n^\pm,i)$ is the corresponding soft factor
for Yang-Mills theory
\begin{equation}
  \label{e:Aqcd}
  \lim_{k_s^\pm\to0}     A_n(\cdots,     a,s^\pm,b,\cdots)=     S^{\rm
    YM}(a,s^\pm,b)\, A_{n-1}(\cdots,a,b,\cdots)\,.
\end{equation}
It depends on
the helicity of the soft gluon~\cite{Berends:1988zn} and is given by
\begin{equation}
  S^{\rm YM}(q,k^\pm, p)={\epsilon^\pm(q,k)\cdot p\over k\cdot p}\,,
\end{equation}
where $\epsilon$ is the polarization vector of the gluon with
momentum $k$. The covariant expression for the soft factor of
the graviton is therefore
\begin{equation}
S^{\mathrm{gravity}}(n^\pm)  =\sum_{i=1}^{n-1}  \,
{\epsilon^\pm(q_L,k_n)\cdot k_i
\, \epsilon^\pm(q_R,k_n)\cdot k_i\over k_n\cdot k_i}\, ,
\label{softgrav2}
\end{equation}
where the graviton polarization tensor $\epsilon^{\pm\pm}$ has
been split into a product of Yang-Mills polarizations
$\epsilon^\pm\otimes\epsilon^\pm$. This expression shows
explicitly that the gravity soft factor in~(\ref{softSM}) is
independent of the choice of reference momenta $q_L$ and $q_R$.
Using
\begin{equation}
  \epsilon^\pm(q_L,k_n)\cdot k_i -
\epsilon^\pm(\tilde q_L,k_n)\cdot k_i=
S^{\rm YM}(q_L,k_n^\pm,\tilde q_L)\, k_n\cdot k_i\,,
\end{equation}
the soft factor of the graviton changes by
\begin{equation}
  \delta S^{\mathrm{gravity}}(n^\pm)  =
S^{\rm YM}(q_L,k_n^\pm,\tilde q_L)\,\sum_{i=1}^{n-1}
\,\epsilon^\pm(q_R,k_n)\cdot k_i=0\,,
\end{equation}
which vanishes by momentum conservation and transversally of
the polarization vectors.

Using standard spinor-helicity notation these soft factors reads (up to
normalization constants)
\begin{align}\label{e:Sqcd}
S^{\rm YM}(a,s^+,b) = \frac{\langle ab\rangle}
{\langle as\rangle \langle sb\rangle}, \qquad
S^{\rm YM}(a,s^-,b) = \frac{\lbrack ab\rbrack}
{\lbrack as\rbrack \lbrack sb\rbrack}\,.
\end{align}

\subsection{A more crossing-symmetric KLT relation from soft limits}
A KLT relation with $(n-2)!^2$ terms for the $n$-graviton
amplitude was recently considered in~\cite{BjerrumBohr:2010ta}.
This expression has a higher degree of manifest crossing
symmetry. Ignoring overall normalization constants, it reads
\begin{align}\label{e:Mzz}
M_n \approx (-1)^n\nonumber \\ &\hspace{-2cm} \frac{ \sum_{\sigma,\gamma}
\widetilde{A}_n(n,\gamma(2,\dots,n-1),1)
\mathcal{S}[ \gamma(2,\dots,n-1)|\sigma(2,\dots,n-1)]_{p_1}
A_n(1,\sigma(2,\dots,n-1),n)}{s_{12\ldots(n-1)}}\,.
\end{align}
On-shell this  expression is  of course ill-defined since then
$s_{12\dots  n-1}=k_n^2=0$. However, the
numerator also vanishes because of the annihilation property~(\ref{monoS}).
In~\cite{BjerrumBohr:2010ta} a prescription for
taking the on-shell limit such that the formula~(\ref{e:Mzz})
gives the correct $n$-point gravity amplitude was provided.
In this section we will demonstrate that the soft limit of gravity
amplitudes imply that the numerator and denominator
indeed vanish at the same rate. This gives an alternative understanding
of why the limit in~(\ref{e:Mzz}) is finite
and corresponds to the proper $n$-point gravity amplitude. Note
that we are using a $\approx$ to remind ourself that the equality is
in terms of a limiting procedure.

We start from the field theory limit of the expression with $(n-3)!^2$
terms in~(\ref{stringPureKLTn})
\begin{align}
\nn M_n ={}& (-1)^{n-3} \sum_{\sigma,\gamma}
\mathcal{S}[\gamma(2,\dots,n-2)|\sigma(2,\dots,n-2)]_{k_1}\\
&\times A_n(1,\sigma(2,\dots,n-2),n-1,n)
\widetilde{A}_n(n-1,n,\gamma(2,\dots,n-2),1)\,.
\label{stringPureKLTFT}
\end{align}
Without loss of generality we can assume that leg $n$ has
positive helicity. Then, using the soft limit $k_n^+\to 0$ for
the color-ordered amplitudes in~(\ref{e:Aqcd}), we get
\begin{align}
\lim_{k_n^+\to0}(-1)^{n-3}M_n \approx& \sum_{\sigma,\gamma} 
S^{\rm YM}(n\!-\!1,n^+,\gamma(2)) S^{\rm YM}(n\!-\!1,n^+,1)
\mathcal{S}[\gamma(2,\dots,n\!-\!2)|\sigma(2,\dots,n\!-\!2)]_{k_1} \nonumber \\
&\hspace{1.5cm}      \times     A_{n\!-\!1}(1,\sigma(2,\dots,n\!-\!2),n\!-\!1)
\widetilde{A}_{n-1}(n\!-\!1,\gamma(2,\dots,n\!-\!2),1)\,.
\end{align}
We note that this limit does not affect the momentum kernel.
One sees that
\begin{eqnarray}
\nn   S^{\rm YM}(n-1,n^+,i) S^{\rm YM}(n-1,n^+,1) &=&
\frac{\langle n-1,1\rangle}{\langle n-1,n\rangle \langle n,1\rangle}
  \frac{\langle   n-1,i\rangle}{\langle  n-1,n\rangle  \langle
    n,i\rangle}\\
&=&{1\over s_{n,n-1}}\, \frac{\langle n-1,1\rangle [n,n-1]}
{\langle n-1,n\rangle \langle n,1\rangle}
  \, \frac{\langle   n-1,i\rangle}{ \langle
    n,i\rangle}\,,
\end{eqnarray}
and in the denominator $s_{n\, n-1}=s_{12\cdots n-2}$. We can
thus rewrite the soft limit of the $n$-point gravity amplitude
as
\begin{align}
&\hspace{-0.6cm}\lim_{k_n^+\to0}(-1)^{n-3}M_n\approx
\frac{\langle n-1,1\rangle\lbrack n,n-1\rbrack}
{\langle n-1,n\rangle \langle n,1\rangle}
\sum_{i=2}^{n-2} \frac{\langle n-1,i\rangle}{
\langle n,i\rangle}\times \\ \sum_{\sigma,\gamma_i}\nonumber
&
\frac{\widetilde{A}_n(n\!-\!1,i,\gamma_i(2,..\,,n\!-\!2),1)
\mathcal{S}[i,\gamma_i(2,..\,,n\!-\!2)|\sigma(2,..\,,n\!-\!2)]_{k_1}
A_n(1,\sigma(2,..\,,n\!-\!2),n\!-\!1)}{s_{12\ldots n\!-\!2}}\,,
\label{midsoftcal}
\end{align}
where the  permutation $\gamma_i$ is over the $n-4$ legs
$\{2,\dots,i-1,i+1,\ldots,n-2\}$ with  $2\leq  i\leq n-2$.

Comparing   with  the   soft   limit  $k_n^+\to0$   of  the
graviton amplitude~\cite{Weinberg:1965nx,Bern:1998sv}
\begin{equation}
  \lim_{k_n^+\to0}M_n(1,2,\dots,n-1,n)  \approx
   -{1\over  \langle  n,1\rangle\langle  n,n-1
    \rangle}\,\sum_{i=2}^{n-2}   {\langle
    n-1,i\rangle\over  \langle
    n,i\rangle}\, \langle 1,i\rangle [i,n]\,
     M_{n-1}(1,2,\dots,n-1)\,,
\end{equation}
we deduce that the terms at the right-hand-side must satisfy
\begin{eqnarray}\label{wKLT}
\nn& &\!\!\!\!\!\! \\ \sum_{\sigma,\gamma_i}  &&\!\!\!\!
\nonumber\frac{\widetilde{A}_n(n\!-\!1,i,\gamma_i(2,..\,,n\!-\!2),1)
\mathcal{S}[i,\gamma_i(2,..\,,n\!-\!2)|\sigma(2,..\,,n\!-\!2)
]_{k_1} A_n(1,\sigma(2,..\,,n\!-\!2),n\!-\!1)}{s_{12\ldots n-2}}\\
&&\hspace{8cm}\approx -\frac{\langle 1i\rangle \lbrack in
\rbrack}{\langle 1,n-1\rangle \lbrack n-1,n\rbrack} M_{n-1}\,,
\end{eqnarray}
Summing  over  $i$  and   using  the  momentum  conservation
identity $\sum_{i=2}^{n-1}\langle 1i\rangle \lbrack  in\rbrack
=0$ one gets the expression~(\ref{e:Mzz})
of~\cite{BjerrumBohr:2010ta} (omitting the overall sign factor)
\begin{align}
M_{n-1} \\ & \!\!\!\!\!\!\!\!\!\!\!\!\!\!\!\!\approx\! \sum_{\sigma,\gamma}
\! \frac{\widetilde{A}_n(n\!-\!1,\gamma(2,\dots,n\!-\!2),1)
\mathcal{S}[\gamma(2,\dots,n\!-\!2)|\sigma(2,\dots,n\!-\!2)]_{k_1}
A_n(1,\sigma(2,\dots,n\!-\!2),n\!-\!1)}{s_{12\ldots n-2}}\,.
\label{crosKLT}
\end{align}
Using that
\begin{align}
\frac{\langle 1i\rangle \lbrack in\rbrack}{\langle 1,n-1
\rangle \lbrack n-1,n\rbrack} = \frac{s_{iq}}{s_{n-1,q}}, \qquad
\mathrm{with} \qquad q \equiv |1\rangle \lbrack n|\,,
\end{align}
it follows from the calculations in \cite{Feng:2010hd}, that
the above soft limit procedure gives an equivalent description
of eq.~\eqref{crosKLT} as the off-shell regularization
introduced in \cite{BjerrumBohr:2010ta} would. Especially, note
that the auxiliary momentum $q$ satisfy all the
requirements~\cite{BjerrumBohr:2010ta,Feng:2010hd} for an
off-shell regularization, \textit{i.e.} $q^2=k_1\cdot q = 0$
and  $q\cdot  k_{n-1}  \neq   0$.

\section{Color and its kinematic factors\label{colordual}}

The usual color  decomposition of tree-level amplitudes  in
gauge theory is given by
\begin{equation}\label{e:YMorig}
 A^{\rm YM}_n =\sum_{\sigma}
 \tr[T^{1}T^{\sigma(2)}\cdots T^{\sigma(n)}]\, A_n(1,\sigma(2),\dots,\sigma(n))\,,
\end{equation}
where $A_n$ are the color-ordered vector amplitudes and $T^i$
generators of the gauge group. The cyclic and the reverse
properties of the trace imply that there are $(n-1)!/2$
independent color factors.

An alternative form of the tree-level gauge amplitude is \cite{DelDuca:1999ha}
\begin{equation}
  \label{e:AnKK}
  A^{\rm YM}_n=  \sum_{\delta}\, c_{1|\delta(2,\dots,n-1)|n} \,
  A_n(1,\delta(2,\dots,n-1),n)\,,
\end{equation}
where the coefficients are products of the structure constants
$(f^a)_{ij}= \tr([T^i,T^j]T^a)$
\begin{equation}\label{e:cgamma}
  c_{\beta_1|\alpha_1\cdots\alpha_{n-1}|\beta_n}=
  f^{\alpha_1}_{\beta_1 i_1} f^{\alpha_2}_{i_1i_3}\cdots
  f^{\alpha_{n-1}}_{j_{n-1}\beta_n}\,.
\end{equation}
An extension of this representation to one-loop order was
provided in~\cite{DelDuca:1999rs}. The equivalence between the
original form~(\ref{e:YMorig}) and~(\ref{e:AnKK}) relies on the
Kleiss-Kuijf relations satisfied by color-ordered amplitudes
~\cite{Kleiss:1988ne}, \textit{e.g.}
\begin{equation}\label{e:photon}
A_n(1,2,3\dots,n)+ A_n(1,3,4,\dots,n,2)+\cdots+
A_n(1,n,2,3,\dots,n-1) =0\,.
\end{equation}
Because of the annihilation property in~\eqref{monoS}, the
color coefficients in~\eqref{e:AnKK} can be arbitrarily shifted
as
\begin{equation}
  \label{e:shiftcKK}
  c_{1|\delta(2,\dots,n-1)|n} \to  c_{1|\delta(2,\dots,n-1)|n} +
\sum_{\sigma}\, g_{\sigma}\,\times\,\cS[
\delta(2,\dots,n-1)|\sigma(2,\dots,n-1)]_{k_n}\,.
\end{equation}
The coefficients $g_\sigma$ can be constants or (not even
necessarily  gauge invariant)  functions  of the  external
polarizations   and  momenta.   Among   the  $(n-2)!$
$g_{\sigma}$-coefficients   $(n-3)!$  are  redundant,   because
they correspond  to shifts in  the kernel  of the  annihilation
property, and only $(n-3)\times (n-3)!$ coefficients are
therefore necessary in~(\ref{e:shiftcKK}).
 The   form  of  the  color  factors
in~(\ref{e:cgamma}) is a specific choice of representation. We
will illustrate this point at the end of this section.

Thanks to the duality between color and kinematic factors,
as noted by Bern-Carrasco-Johansson~\cite{Bern:2008qj},
one expects that the gravity tree-level amplitudes take a
similar form~\cite{Bern:2010yg,KiermaierTalk}
\begin{equation}
  \label{e:AnKKgrav}
  M_n= (-1)^{n-3}\, \sum_{\sigma}\, n_{1|\sigma(2)\cdots\sigma(n-1)|n} \,
  A_n(1,\sigma(2,\dots,n-1),n)\,.
\end{equation}
To derive such a representation for the gravity amplitude we
will define  $(i^0_2,\dots,i^0_{n-1})=(2,3,4,\dots,n-1)$  to   be
the identity, $(i^1_2,\dots,i^1_{n-1})=(n-1,2,3,\dots,n-2)$ the
cyclic right        permutation       of        order $n-1$,
and $(i^k_2,\cdots,i^k_{n-1})=(n-k,n-k+1\cdots,n-1-k)$ the
$k$th iteration of  the cyclic right  permutation (the
notation is to be  understood modulo $n-2$ as $i^2_{n-1}=2$).

We start by considering the symmetrized version of the
$n$-graviton amplitude
\begin{equation}
  M_n(1,2,\dots,n)=    {1\over    n-2}\,    \sum_{k=0}^{n-3}
  M_n(1,i^k_2,\dots,i^k_{n-1},n)\,.
\end{equation}
One can consider a more general averaging with non-equal weight
factors. Plugging in the $(n-3)!^2$
formula~(\ref{stringPureKLTn}) for the $n$-point graviton
amplitude in  this expression  we  get 
\begin{equation}
  \label{eq:Mn-2}
\begin{split}
   M_n={}& {(-1)^{n-3}\over
  n-2}\,\sum_{j=0}^{n-3}\sum_{\sigma,\gamma}
  \,\cS[\gamma(i^j_2,\dots,i^j_{n-2})|\sigma(i^j_2,
  \dots,i^j_{n-2})]_{k_1}\cr
&\times
   A_n(1,\sigma(i^j_2,\dots,i^j_{n-2}),i^j_{n-1},n)\,
   \widetilde{A}_n(1,\gamma(i^j_2,\dots,i^j_{n-2}),n,i^j_{n-1})\,.
\end{split}\end{equation}
The same
  manipulations can  be done  starting from the  $j$-dependent formula
  in~(\ref{eq:Mnfinal}). The equivalence  of the resulting expressions
  is obtained by a repeated use of the monodromy
  relations~\cite{BjerrumBohr:2009rd,BjerrumBohr:2010zs} between the
color-ordered amplitudes. Defining the coefficients
\begin{equation}\begin{split}
  n_{1|\sigma(i^j_2,\dots,i^j_{n-2})i^j_{n-1}|n}
  \\ &\hspace{-3cm} \equiv {1\over n-2}\,
\sum_{\gamma}\,\cS[\gamma(i^j_2,\dots,i^j_{n-2})|
\sigma(i^j_2,\dots,i^j_{n-2})]_{k_1}\times
   \widetilde{A}_n(1,\gamma(i^j_2,\dots,
   i^j_{n-2}),n,i^j_{n-1})\,,
\end{split}\end{equation}
and remarking that the sum over all $\sigma$-permutations
composed with the cyclic permutations $i^k$ provide all
permutations of $\{2,\dots,n-1\}$, we can
rewrite~\eqref{eq:Mn-2} as

\begin{equation}
  \label{eq:Mn-22}
\begin{split}
& M_n=(-1)^{n-3}\, \sum_{\delta}\, n_{1|\delta(2,\dots,n-1)|n}\,
 A_n(1,\delta(2,\dots,n-1),n)\,,
\end{split}\end{equation}
with
\begin{equation}
  \label{e:Defn}\begin{split}
 n_{1|\delta(2,\dots,n-1)|n} ={}& {1\over n-2}\,
\sum_{\gamma}\,\cS[\gamma(\delta(2),
\dots,\delta(n-2))|\delta(2),\dots,\delta(n-2)]_{k_1}\times\cr
&\times   \widetilde{A}_n(1,\gamma(\delta(2),\dots,
\delta(n-2)),n,\delta(n-1))\,.
\end{split}\end{equation}
This provides the form of the $n$-point graviton amplitude
similar to~(\ref{e:AnKK}), where the color factors have been
replaced by kinematic factors.

We   remark   that   the  manipulations   leading
to~(\ref{eq:Mn-22}) and~(\ref{e:Defn}) are valid in string
theory as well,  with the  replacement of the  field theory
$\cS$  by its stringy  version  $\cS_{\alpha'}$   and  the
field theory  amplitudes $A_n$  and $\widetilde{A}_n$ by  the
correspond string amplitudes.

There are $(n-2)!$ color and kinematic coefficients in the
decomposition in~(\ref{e:AnKK}) and~(\ref{e:AnKKgrav}) which is
the same as the number of independent numerator factors in the
BCJ decomposition of amplitudes in terms       of      cubic
vertices~\cite{Bern:2008qj,Bern:2010yg,KiermaierTalk}. The
expression~(\ref{e:Defn})  provides a possible constructive
definition  for these independent numerator factors.

\medskip

We illustrate the construction  of the color coefficient
in~(\ref{e:AnKK}) and the kinematic 
coefficients~(\ref{e:AnKKgrav}) in the case of the four points.
By applying~(\ref{e:Defn}) we have
\begin{equation}
  \hat     c_{1|23|4}=    {1\over2}     \,     \cS[2|2]_{k_1}    \,
  A^{s}_4(1243), \qquad
  \hat c_{1|32|4}= {1\over2} \, \cS[3|3]_{k_1} \, A^{s}_4(1342)\,,
\end{equation}
where the scalar amplitudes are defined by~\cite{Bern:2008qj}
\begin{equation}
 A^{s}_4(1243)= A^{s}_4(1342)={c_s\over s}+{c_u\over u}\,,
\end{equation}
with  the  numerator  factors  given by
$c_s= -\sum_i f^i_{12}f^i_{34}$  and
$c_u=\sum_i f^i_{31}f^i_{24}$.
Therefore we get
\begin{equation}
 \hat c_{1|23|4}= {1\over2} \, ( c_s+{s\over u}\,c_u), \qquad
  \hat c_{1|32|4}= {1\over2} \, (c_u+{u\over s}c_s)\,.
\end{equation}
These coefficients  have a non-local momentum dependent  form which can
be  removed by  using  the freedom  in~(\ref{e:shiftcKK}). We can, for instance, shift these
coefficients, without modifying the amplitude, by
\begin{equation}\label{e:shift}
   c_{1|\sigma(23)|4}=\hat c_{1|\sigma(23)|4}
   +{1\over2}\left( {c_u\over su}- {c_s\over s^2}\right) \,
  \cS[\sigma(2,3)|3,2]_{k_4}\,,
\end{equation}
to obtain the form of the coefficients in~(\ref{e:cgamma})
\begin{equation}
 c_{1|23|4}= c_s, \qquad
   c_{1|32|4}= c_u\,.
\end{equation}
The same  manipulations can  be done for  the kinematic
coefficient when applying~(\ref{e:Defn}) with the vector
amplitudes
\begin{equation}
 A_4(1243)={n_s\over s}+{n_u\over u}, \qquad A_4(1234)=
 -{n_s\over s}+{n_t\over t}, \qquad A_4(1324)=
 -{n_u\over u}-{n_t\over t}\,.
\end{equation}
Exchanging  the  vector and  the  scalar  amplitudes  in these
formulaes exchange  the color  and  kinematic coefficients.

\section{Conclusions\label{conclu}}
This paper has  been devoted  to the  properties of  the
momentum kernel that lies behind closed-open string
factorization. This  kernel is  the  central object in the
relation between gravity and gauge theory amplitudes in both
string theory and field theory.  We have shown  that this
kernel defines a set of linear equations that annihilate
color-ordered Yang-Mills amplitudes: it is the generator of
monodromy relations between these amplitudes. The kernel has
also an interesting connection relating to kinematic
factors.

Although no natural origin of this momentum kernel is known
from the structure of  the (effective) gravity and gauge theory
Lagrangians, we have shown that this  object arises  naturally
when one constructs  closed (type~II or heterotic) string
amplitudes.  This parallels (and complements) the $S$-matrix
based motivation for this kernel in~\cite{BjerrumBohr:2010ta}.
The very simple product form of this kernel makes it easy to
implement the   KLT-form  of   the   amplitudes
in~(\ref{eq:Mnfinal}),   the annihilation  relations
in~(\ref{monoS}), and  the  definition of  the kinematic  factors
in~(\ref{e:Defn}).

Our string-based derivation of the annihilation property and,
consequently,  the   monodromy  relations  between
color-ordered amplitudes did not use any  detailed properties
of the spectrum of the theory. The momentum kernel follows in a
universal way from the phases of the operator product
expansion. We therefore expect that such a kernel will enter in
relations between ordered correlators in other contexts that
are based on a conformal field theory, as in solid state
physics. It would be interesting to have this possible
application elucidated in greater detail.

\section*{Acknowledgements}

P.V. would like to thank the Niels Bohr International Academy,
where part of this work was carried out, for hospitality. This
research has been supported in part by the French Embassy of
Denmark.

\end{document}